\documentstyle[aasms,psfig]{article}
\begin{document}

\title{Gravitational Lensing by Power-Law Mass Distributions: 
A Fast and Exact Series Approach}
\author{Kyu-Hyun Chae\footnote{Department of Physics \& Astronomy, 
University of Pittsburgh, Pittsburgh, PA 15260}, 
Valery K. Khersonsky$^1$, and David A. Turnshek$^1$}

\begin{abstract}
We present an analytical formulation of gravitational lensing using 
familiar triaxial power-law mass distributions, where the 3-dimensional 
mass density is given by $\rho(X,Y,Z) = \rho_0 \left[1 + (\frac{X}{a})^2 + 
(\frac{Y}{b})^2 + (\frac{Z}{c})^2 \right]^{-\nu/2}$. 
The deflection angle and  magnification factor are obtained 
analytically as Fourier series.
We give the exact expressions for the deflection angle and magnification 
factor. The formulae for the deflection angle and magnification factor
given in this paper will be useful for numerical studies of observed lens
systems. An application of our results to the Einstein Cross can be found in 
Chae, Turnshek, \& Khersonsky (1998). Our series approach can be viewed as a
user-friendly and efficient method to calculate lensing properties that is 
better than the more conventional approaches, e.g., numerical integrations, 
multipole expansions.
\end{abstract}


\section{Introduction}
Gravitational lenses in nature usually consist of a primary lensing 
galaxy and, in many cases, perturbation field caused by, for example,
nearby galaxies and clusters of galaxies along the sight-line
 (see Keeton, Kochanek, \& Seljak 1997).
Thus, modeling an observed lens requires the construction of a realistic
mass model, including accounting for any perturbation that may be present.
This paper developes a (semi-)analytical method to calculate lensing due 
to the primary mass distribution.

Lensing objects (e.g.\ galaxies or galaxy clusters) show, 
in general, 3-dimensional mass distributions which are non-spherically 
symmetric. Elliptical galaxies are known to be triaxial 
in shape, as well as oblate and prolate (see, e.g., Ryden 1992).
Spiral galaxies can be considered flattened spheroids.
Clusters of galaxies are also known to be non-spherical in shape.
Gravitational lensing by these concentrations of mass 
in the universe is difficult to deal with mathematically.
For a projected surface density of elliptical shape Schramm (1990) 
introduced a relatively simple formulation 
equivalent to the early formulation 
by Bourassa, Kantowski, \& Norton (1973) and Bourassa \& Kantowski 
(1975, see also Bray 1984).
In the above formulations the deflections are given as integrals.
Thus, one can deal with lensing by elliptical mass distributions
through numerical integrations. While numerical integration is a
way of dealing with lensing by elliptical mass distributions, 
it has not been preferred by most researchers.

To ``simulate'' real lenses in nature while avoiding mathematical
complexities or costly numerical integrations, 
several lens models have been considered in the literature. 
First, circularly symmetric lenses perturbed by an external shear term 
(also called a quadrupole term) were used
(e.g.\ Chang \& Refsdal 1979, 1984; Kovner 1987a; 
Kochanek 1991; Wambsganss \& Paczy\'{n}ski 1994).
In addition, Kochanek (1991) considered  other 
types of perturbations (internal \& mixed). 
This approach has the great advantage of mathematical simplicity.
 It could be a good approximation of a real lens in some situations. 
However, this approach is somewhat artificial and does not take into account
the real distributions of mass.
 Secondly, lenses with elliptical deflection potentials were studied 
(e.g.\ Kovner 1987b,c; Blandford \& Kochanek 1987a; 
Kochanek \& Blandford 1987; Kochanek et al.\ 1989). 
Elliptical potentials are easier to handle than elliptical mass distributions.
For a small ellipticity elliptical potentials correspond to physical 
(elliptical) mass distributions via the Poisson equation, 
but for a larger ellipticity (e.g.\ $0.5 \lesssim \epsilon$) 
the corresponding mass distributions obtain unphysical 
dumbbell shapes (see Kassiola \& Kovner 1993).
Thirdly, a multipole expansion of elliptical mass distributions 
was considered by Schneider \& Weiss (1991) who 
expanded a 2-dimensional distribution of mass with elliptical symmetry 
and calculated the corresponding potential coefficients. 
Thus, in this approach a (nearly) elliptical mass distribution  
is described by the several lowest-order terms in the expansion.
Lastly, for a few special cases the deflection angles
were obtained in simple (closed) forms. 
For elliptical ``isothermal'' lenses (where the 
surface mass density $\Sigma \sim r^{-(\nu-1)}$ [$\nu = 2$] at large $r$), 
the deflection angles were calculated by Kassiola \& Kovner (1993) 
and Kormann, Schneider, \& Bartelmann (1994). Kassiola \& Kovner (1993) also
calculated the deflection angles 
for other integer values of $\nu$, i.e.\ $\nu$ = 3, 4, 5, etc. 
For singular power-law mass distributions, 
the deflection angle was obtained by Grogin \& Narayan (1996).

All of the above lens models avoid numerical integrations and are relatively 
easy to use. However, we know that most (if not all) of them are limited
in their applications.
For the generalized non-circular power-law mass distributions where the 
parameter $\nu$ and core radius are arbitrary, numerical integrations 
(Bourassa \& Kantowski 1975; Schramm 1990) and multipole expansions 
(Schneider \& Weiss 1991) have been considered and the former approach 
has sometimes been used to calculate lensing properties (e.g. most recently
by Keeton and Kochanek 1997). In the Schneider \& Weiss (1991)
approach an elliptical mass distribution is expanded and
several lowest-order terms are used to describe the original mass 
distribution. This approximation can be effective for a less elliptical mass
distribution because only a few terms suffice to describe the original
mass distribution. However, for a highly elliptical mass distribution
this approach becomes increasingly less effective because many terms are 
necessary to approximate the original mass distribution. Also, we do not,
at present, find any examples in the literature where this approach was used.
For the above reasons this approach will not be further discussed.
In this paper we present an analytical approach to lensing by the generalized 
power-law mass distributions. In our approach all of the terms in the
infinite series can be integrated in closed forms 
(as hypergeometric functions), and the series themselves
have well-defined convergence properties. 
As a result, we obtain the deflection and magnification in 
ready-to-use forms. Our approach is a straightforward and user-friendly way to 
calculate lensing properties more effectively than the more conventional 
methods mentioned above. 
Our results can be used by those who want to use (exactly) elliptical 
surface mass densities (or, 3-dimensional triaxial mass distributions directly)
and who do not want to use numerical integrations.
We applied our results to the Einstein Cross and they were very
effective (Chae, Turnshek, \& Khersonsky 1998).
Our approach is as follows.

We represent the deflection potential as a Fourier series 
for an arbitrary projected surface density. From the deflection potential 
we obtain the deflection angle as a series. We calculate the projected 
surface mass density of the triaxial power-law mass distribution and thus 
relate the parameters of the surface mass density to those of the 3-dimensional
mass density. We calculate the coefficients in the series of the deflection 
angle for the surface density. We obtain expressions for the deflection angle 
and magnification factor.

The above procedures are described in \S 2; the details of mathematics 
can be found in the Appendices. In \S 3 we 
describe how the behaviors of critical curves and caustics depend on 
the parameters of the elliptical power-law mass distributions.
Concluding remarks are given in \S 4.

\section{Mathematical Formulation}

\subsection{{\it Basic Equations, Definitions and Notation}}

In this subsection we briefly summarize the basic equations of gravitational 
lensing and define the coefficients of the series of the deflection potential.
Our treatment of lensing is based on the scalar potential formalism
given by Schneider (1985) and Blandford \& Narayan (1986). 
Our notation is similar to that used by 
Schneider, Ehlers, \& Falco (1992, hereafter SEF) 
and Blandford \& Kochanek (1987b).

When a light ray emitted by a quasar (i.e.\ the source) 
at angular diameter distance $D_{s}$ passes through a transparent
distribution of mass (i.e.\ the deflector) at $D_{d}$, 
the light ray experiences a deflection and the
condition for the deflected light ray to reach 
the observer is given by the lens equation (SEF)
\begin{equation}
 \vec{\eta} = \frac{D_{s}}{D_{d}} \vec{\xi} - D_{ds} \hat{\alpha}(\vec{\xi}).
\end{equation}
Here $\vec{\eta}$ is the position vector of the quasar on the source plane and 
$\vec{\xi}$ is the impact vector of the light ray on the lens plane, 
both with respect to the optical axis (defined below). 
The parameter $\hat{\alpha}(\vec{\xi})$ is the deflection angle 
and $D_{d}, D_{s}$ and $D_{ds}$ 
are the angular diameter distances to the deflector, 
the source, and from the deflector to the source, respectively.
The optical axis is defined as the infinite line joining 
the observer and the center of mass of the lens.
By introducing an arbitrary length scale $\xi_{0}$ and 
defining $\vec{x} = \vec{\xi}/\xi_0$ (deflector vector),
$\vec{x}_{s} = \vec{\eta}/(\xi_{0} D_{s}/D_{d})$ (source vector), 
the above equation becomes the dimensionless lens equation
\begin{equation}
\vec{x}_{s} = \vec{x} - \vec{\alpha}(\vec{x}),
\end{equation}
where $\vec{\alpha}(\vec{x})$ is the scaled deflection angle which is
related to the deflection potential $\psi(\vec{x})$ 
by $\vec{\alpha}(\vec{x}) = \nabla \psi(\vec{x})$. 
The deflection potential is given in polar coordinates (SEF) by 
\begin{equation}
\psi(r,\phi) = \frac{1}{\pi} \int_0^{2\pi} d \phi'\int_0^{\infty}
\kappa(r',\phi') \
\ln [r^{2} + r'^{2} - 2rr'\cos(\phi - \phi')]^{1/2} r' dr',
\end{equation}
where the dimensionless surface density $\kappa (\vec{x})$ 
is obtained by dividing the surface density
 $\Sigma(\vec{\xi})$ by $\Sigma_{cr}$, i.e.\ 
$\kappa (\vec{x}) =\Sigma (\xi_{0} \vec{x}) / \Sigma_{cr}$,
and the critical surface mass density $\Sigma_{cr}$ is defined by
$\Sigma_{cr} = (4 \pi G /c^{2})^{-1} ( D_{d} D_{ds}/D_s)^{-1}$.

By expanding the logarithmic function in equation (3) 
for $r' > r$ and $r' < r$ respectively,
 the deflection potential can be written as
\begin{eqnarray}
\psi (r,\phi) & = & \frac{A_{0}(r)}{\pi} \ln r - 
\sum_{n=1}^{\infty} \frac{\cos n \phi}{n \pi} 
\left[\frac{B_{n}(r)}{r^{n}} + r^{n} D_{n}(r)\right] \nonumber \\
   &  & +\frac{F(r)}{\pi} - \sum_{n=1}^{\infty} \frac{\sin n \phi}{n \pi}
    \left[\frac{C_{n}(r)}{r^{n}} + r^{n} E_{n}(r)\right],
\end{eqnarray}
and the deflection angle components are
\begin{eqnarray}
\alpha_{r} & =  & \frac{1}{\pi r} A_{0}(r) \nonumber \\
  &   & - \frac{1}{\pi r} \sum_{n=1}^{\infty} \left\{ 
\cos (n \phi) \left[-\frac{B_{n} (r)}{r^{n}} + 
        r^{n} D_{n}(r)\right] + \sin (n \phi) 
\left[-\frac{C_{n} (r)}{r^{n}} + r^{n} E_{n} (r)\right] \right\}, \\
\alpha_{\phi} & = & \frac{1}{\pi r} \sum_{n=1}^{\infty} 
\left\{ \sin (n \phi) \left[\frac{B_{n} (r)}{r^{n}} + r^{n} D_{n} (r)\right] - 
 \cos (n \phi) \left[\frac{C_{n} (r)}{r^{n}} + r^{n} E_{n} (r)\right] \right\},
\end{eqnarray} 
where the coefficients of the series are defined as follows:
\begin{eqnarray}
A_{0}(r) & = & \int_{0}^{2 \pi} d \phi' \int_{0}^{r} 
              \kappa (r',\phi') r' dr',  \\
B_{n}(r) & = & \int_{0}^{2 \pi} d \phi' \cos n \phi' 
           \int_{0}^{r} \kappa (r',\phi') r'^{n+1}  dr',  \\
C_{n}(r) & = & \int_{0}^{2 \pi} d \phi' \sin n \phi' 
           \int_{0}^{r} \kappa (r',\phi') r'^{n+1}  dr',  \\
D_{n}(r) & = & \int_{0}^{2 \pi} d \phi' \cos n \phi' 
          \int_{r}^{\infty} \kappa (r',\phi') \frac{1}{r'^{n-1}} dr',  \\
E_{n}(r) & = & \int_{0}^{2 \pi} d \phi' \sin n \phi' 
           \int_{r}^{\infty} \kappa (r',\phi') \frac{1}{r'^{n-1}} dr',
\end{eqnarray}
and
\begin{equation}
F(r)  = \int_{0}^{2 \pi} d \phi' \int_{r}^{\infty}  
        \kappa (r',\phi') \ln r' \cdot r' dr'.
\end{equation}
 
From equation (2) the image magnification factor is given by
\begin{equation}
{\cal M} =  \left[ \det\left|\frac{\partial 
       \vec{x}_{s}}{\partial \vec{x}}\right| \right]^{-1},
\end{equation}
where $\left( \frac{\partial \vec{x}_s}{\partial \vec{x}} \right)$ 
is the Jacobian matrix in $\vec{x}_s = \vec{x}_s (\vec{x})$.  
A point source at $\vec{x}_s$ on the source plane forms an image 
at $\vec{x}$ on the lens plane magnified by a factor of $|{\cal M}|$.

\subsection{{\it The Projected Surface Mass Density}}
The 3-dimensional triaxial power-law distribution of mass is described by
\begin{equation}
\rho (X,Y,Z) = \rho_{0} \left[1 + \left(\frac{X}{a}\right)^{2} + 
\left(\frac{Y}{b}\right)^{2} + \left(\frac{Z}{c}\right)^{2} \right]^{-\nu / 2},
\end{equation}
where ($X, Y, Z$) are the body coordinates attached to the lensing object.
 The positive constants ($a, b, c$) represent the core sizes along each axis.
In passing, we mention that a more conventional parameterization 
is obtained through $b=au$ and $c=av$, where $1 \geq u \geq v $ by convention.
The constant $\rho_{0}$ is the density at the center.
The radial index $\nu$ determines how the mass density decreases at large $r$. 
If $\nu > 3$, the lens has a finite total mass ($M$) 
and $\rho_0$ is related to the total mass by 
$\rho_0 = M [2\pi abc \mbox{B} (3/2,\mu)]^{-1}$, 
where $\mu \equiv (\nu-3)/2$ and B($a$,$b$) is the Beta function. 
When $\nu = 2$, the mass distribution is called ``isothermal'' 
since the mass density decreases like 
the singular isothermal sphere at large $r$.
 A physical distribution of mass must satisfy  
$\nu > 1$ because  $\nu = 1$ corresponds
to a constant  surface density on the lens plane (see eq.\ [15]).

Since a lensing galaxy (or any lensing object) can be oriented 
in an arbitrary direction relative to the lens plane, 
three parameters are needed to relate the body coordinates $(X, Y, Z)$ 
to the lens coordinates ($x, y, z$), 
where $z$ is the direction toward the observer. 
These three parameters are the Eulerian angles ($\alpha, \beta$, $\gamma$) 
(see, e.g., Goldstein 1980). 
We have $X_i = \sum_j a_{ij} x_j$, where $X_i = (X, Y, Z)$, $x_j = (x, y, z)$,
 and the transformation matrix ($a_{ij}$) can be found 
in Goldstein (1980, eq.\ [4-47]) with $i,j = 1, 2, 3$.
Using this transformation we find
\[ 1 + \frac{X^{2}}{a^{2}}+\frac{Y^{2}}{b^{2}}+\frac{Z^{2}}{c^{2}} 
      = A z^{2} + 2 B(x,y) z + C(x,y),  \]
where $A = a_{13}^{2}/a^{2} +a_{23}^{2}/b^{2} +a_{33}^{2}/c^{2}$ and
\begin{eqnarray*}
 B(x,y) & = & B_{1} x + B_{2} y \hspace{0.02in}; 
 \hspace{0.1 in} B_{1} = \frac{a_{11} a_{13}}{a^{2}} + 
 \frac{a_{21} a_{23}}{b^{2}}
 + \frac{a_{31} a_{33}}{c^{2}},\hspace{0.1 in} B_{2} = 
 \frac{a_{12} a_{13}}{a^{2}} + \frac{a_{22} a_{23}}{b^{2}} 
 + \frac{a_{32} a_{33}}{c^{2}},  \\   
C(x,y) & = & 1 + C_{1} x^{2} + C_{2} y^{2} + 2 C_{3} x y \hspace{0.02in}; 
\hspace{0.1 in}C_{1} = \frac{a_{11}^{2}}{a^{2}} 
+\frac{a_{21}^{2}}{b^{2}} +\frac{a_{31}^{2}}{c^{2}},  
\hspace{0.1 in} C_{2} = \frac{a_{12}^{2}}{a^{2}} 
+\frac{a_{22}^{2}}{b^{2}} +\frac{a_{32}^{2}}{c^{2}},\\ 
 &  &  \hspace{2.0 in}C_{3} = \frac{a_{11} a_{12}}{a^{2}} 
+ \frac{a_{21} a_{22}}{b^{2}} + \frac{a_{31} a_{32}}{c^{2}}.
\end{eqnarray*}

Now the 2-dimensional surface density on the lens plane can be evaluated as
\begin{eqnarray}
\Sigma (x,y) & = & \int_{-\infty}^{\infty} \rho(x,y,z) dz  \nonumber  \\
     & =  & 2^{2\mu +1} \mbox{B}(\mu +1,\mu +1) 
      \frac{\rho_{0}}{\sqrt{A}} \frac{1}{[f(x,y)]^{\mu + 1}},
\end{eqnarray}
where  $f(x,y) \equiv C(x,y) -  [B(x,y)]^{2}/A$. 
Note that $\mu [\equiv (\nu-3)/2] > -1$.
\newline  
We rewrite $f(x,y)$ as
\begin{equation}
f(x,y)  =  1 + c_{x} x^{2} + c_{y} y^{2} + c_{xy} xy , 
\end{equation}
where $c_{x} = C_{1} - B_{1}^{2}/A, c_{y} = C_{2} - B_{2}^{2}/A$, 
      and $c_{xy} = 2(C_{3} - B_{1} B_{2}/A)$. 
In polar coordinates equation (16) becomes
\begin{equation}
f(r,\phi) = 1 + r^{2} [P + Q \sin (2 \phi + S)],
\end{equation}
where $ P = (c_{x} + c_{y})/2 $, $ S = \tan^{-1} [(c_{x} - c_{y})/c_{xy}]$, 
   and $ Q = (c_{x} - c_{y})/ (2 \sin S)$.

Equations (15) and (17) show that the projected surface 
density of a triaxial mass distribution (eq.\ [14]) 
is described by a set of concentric ellipses with constant ellipticity
(see, e.g., Stark 1977). 
The ellipticity of an ellipse, $\epsilon$, is given by
\begin{equation}
\epsilon = 1 - \frac{r_{min}}{r_{max}} = 1 - 
\sqrt{\frac{1 - |e|}{1 + |e|}},
\end{equation}
where  $r_{min}$ and $r_{max}$ are the semi-minor and 
semi-major axes of an ellipse, respectively, and $e \equiv Q/P$.  
The ratio $|e|$ is a measure of the shape of an ellipse 
and $ 0 \leq |e| < 1$.

The dimensionless surface density as a function of the deflector vector is 
given by
$$ \kappa (\vec{x}) = \kappa (r,\phi) = \frac{\kappa_{0}}
       {\{1 + r^{2} [P + Q \sin (2 \phi + S)]\}^{\mu + 1}}. \eqno(19a) $$
Equation (19a) can be rewritten in the following more convenient form
$$
\kappa(r,\phi) = \kappa_0 \left\{ 1 + \left(\frac{r}{r_c}\right)^2
[1 + e \cos 2(\phi-\phi_0)] \right\}^{-(\mu+1)}, \eqno(19b) $$
where the ``core radius'' $r_c \equiv 1/\sqrt{P}$ and the orientation angle
$\phi_0 \equiv \frac{\pi}{4} - \frac{S}{2}$, which becomes the standard
position angle (P.A., north through east) if $e > 0$ (see also
Keeton \& Kochanek 1997). In  equations (19a,b)
\setcounter{equation}{19}
\begin{equation}
\kappa_{0} = \frac{2^{2\mu +1} 
 \mbox{B}(\mu +1,\mu +1)\rho_{0}}{\sqrt{A} \cdot \Sigma_{cr}}.
\end{equation}

\subsection{{\it The Deflection Angle and Magnification Factor}}
The coefficients of the deflection angle (equations [7] - [11]) 
are calculated in Appendix A.
Using the results of Appendix A, we are now ready to  
consider the full expressions for the deflection angle  and  magnification 
factor at
 $\vec{x}$ on the lens plane due to 
the mass  distribution given by equation (14) or equation (19).
From Appendix A, all odd-numbered coefficients vanish 
and the even-numbered coefficients can be written as follows 
by introducing three functions of $r$ [$I^{(0)}(r)$, $I^{(1)}_{2m}(r)$, 
and $I^{(2)}_{2m}(r)$]: 
\begin{equation}
A_0(r) = \kappa_0 \pi r I^{(0)}(r),
\end{equation}
\begin{equation}
\left\{ \begin{array} {l} B_{2m}(r) =  
      \kappa_0 \pi  r^{1+2m} I_{2m}^{(1)}(r) \cos(m\delta)\\
 C_{2m}(r) =  \kappa_0 \pi r^{1+2m} I_{2m}^{(1)}(r)\sin(m\delta) 
\end{array} \right.,
\end{equation}
and
\begin{equation}
\left\{ \begin{array} {l} D_{2m}(r) =  
    \kappa_0 \pi r^{1-2m} I_{2m}^{(2)}(r) \cos(m\delta)\\
E_{2m}(r) =  \kappa_0 \pi r^{1-2m} I_{2m}^{(2)}(r)\sin (m\delta) 
\end{array} \right.,
\end{equation}
where $\delta \equiv \pi/2 - S$. 

Now the deflection angle components, equations (5) and (6), 
 can be written in the following simple forms
\begin{eqnarray}
\alpha_r & = & \kappa_0 I^{(0)}(r) + \kappa_0 \sum_{m=1}^{\infty} 
\left[I^{(1)}_{2m}(r)-I^{(2)}_{2m}(r)\right] \cos [m(2\phi-\delta)], \\
\alpha_{\phi} & = & \kappa_0 \sum_{m=1}^{\infty} 
\left[ I^{(1)}_{2m}(r)+I^{(2)}_{2m}(r)\right] \sin [m(2\phi-\delta)].
\end{eqnarray}  
Using the results of Appendix A, $I^{(0)}(r), I^{(1)}_{2m}(r)$, 
and $I^{(2)}_{2m}(r)$ can be written as follows.
\newline
For $r \leq 1/\sqrt{P-|Q|}$
$$ I^{(0)}(r) = h(r)[\varepsilon_2(r)]^{\mu} \sum_{k=0}^{\infty} 
\sum_{l=0}^{k} (-1)^l\zeta_1(k,l,0,\mu) 
[\varepsilon_2(r)]^{l} {_2F_1}(-l-\mu,-l-\mu;1;\varepsilon_1(r)),  
\eqno(26a) $$
for $r > 1/ \sqrt{P-|Q|}$ and  $\mu \neq 0$
$$ I^{(0)}(r) = \frac{1}{\mu}\left[ \frac{1}{r\sqrt{P^2-Q^2}} 
-h(r)[\varepsilon_2(r)]^{\mu}\sum_{k=0}^{\infty}
 [\varepsilon_2(r)]^{k}{_2F_1}(-k-\mu,-k-\mu;1;\varepsilon_1(r)) 
\right], \eqno(26b) $$
and for  $\mu = 0$ 
$$ I^{(0)}(r) = \frac{1}{r\sqrt{P^2-Q^2}} \ln \left[\frac{\sqrt{P^2-Q^2} 
\sqrt{(P^2-Q^2)r^4 +2Pr^2 +1} +(P^2-Q^2)r^2 +P} 
{\sqrt{P^2-Q^2}+ P} \right].  \eqno(26c) $$
\newline
For  $r \leq 1/\sqrt{P-|Q|}$
\begin{eqnarray*}
 I_{2m}^{(1)}(r) & = &
  \left(-\frac{Q}{|Q|}\right)^m h(r) 
    [\varepsilon_1(r)]^{\frac{m}{2}} [\varepsilon_2(r)]^{\mu} \\
 &  &  \times \sum_{k=0}^{\infty}\sum_{l=0}^{k}(-1)^l
     \zeta_1(k,l,m,\mu) [\varepsilon_2(r)]^{l}
 {_2F_1}(m-l-\mu,-l-\mu;m+1;\varepsilon_1(r)),   
\end{eqnarray*} 
$$ \eqno(27a) $$
and for $r> 1/\sqrt{P-|Q|}$, if  $\mu$ is not integer,
\begin{eqnarray*}
 I_{2m}^{(1)}(r) & = & -\left(-\frac{Q}{|Q|}\right)^m h(r) 
     [\varepsilon_1(r)]^{\frac{m}{2}}
    \frac{\Gamma(\mu-m)}{\Gamma(\mu+1)\Gamma(m+1)} \\
 &  & \times \left\{ [\varepsilon_2(r)]^{\mu} \sum_{k=0}^{\infty} 
   \frac{\Gamma(k+m+\mu+1)}{\Gamma(k-m+\mu+1)}
[\varepsilon_2(r)]^k {_2F_1}(m-k-\mu,-k-\mu;m+1;\varepsilon_1(r)) \right. \\
&  & \hspace{0.25in} -\left. [\varepsilon_2(r)]^m \sum_{k=0}^{\infty} 
   \frac{\Gamma(k+2m+1)}{\Gamma(k+1)}
 [\varepsilon_2(r)]^k {_2F_1}(-k,-k-m;m+1;\varepsilon_1(r)) \right\}.
\end{eqnarray*}
$$  \eqno(27b) $$
\setcounter{equation}{27}
And for any $r > 0$
\begin{eqnarray} 
 I_{2m}^{(2)}(r) & = & \left(-\frac{Q}{|Q|}\right)^m h(r) 
  [\varepsilon_1(r)]^{\frac{m}{2}} 
[\varepsilon_2(r)]^{\mu} \zeta_2(m,\mu) \nonumber \\
           &  & \times \sum_{k=0}^{\infty} [\varepsilon_2(r)]^{k}
{_2F_1}(m-k-\mu,-k-\mu;m+1;\varepsilon_1(r)).
\end{eqnarray}
In equations (26), (27), and (28) we defined the following:
\begin{eqnarray}
h(r)&=&\frac{r}{\sqrt{(1+Pr^2)^2-(Qr^2)^2}}, \\
\varepsilon_1(r)& =& \left[1-\sqrt{1- \left( \frac{Qr^2}{1+Pr^2} 
\right)^2}\right] 
\left[1+\sqrt{1- \left( \frac{Qr^2}{1+Pr^2} \right)^2}\right]^{-1}, \\
\varepsilon_2(r)& = &\frac{1}{2} \left[ \frac{1+Pr^2}{(1+Pr^2)^2-(Qr^2)^2} +
 \frac{1}{\sqrt{(1+Pr^2)^2-(Qr^2)^2}} \right], \\
\zeta_1(k,l,m,\mu)&=&\frac{\Gamma(k+1)\Gamma(k+\mu+1)\Gamma(m+l+\mu+1)} 
{\Gamma(l+1)\Gamma(k-l+1)\Gamma(\mu+1)\Gamma(l+\mu+1)\Gamma(m+k+2)}, 
\end{eqnarray}
and
\begin{equation}
\zeta_2(m,\mu) = \frac{\Gamma(m+\mu)}{\Gamma(m+1)\Gamma(\mu+1)}.
\end{equation}
Note that $0 \leq \varepsilon_1(r) < 1$ and $ 0 < \varepsilon_2(r) \leq 1$ 
 for $ 0 \leq r < \infty$.
In equations (26) and (27), the range of $r$ for each expression corresponds
to the range for which the expression converges rapidly. 
Outside this range each expression will still converge but slowly. 
Equation (28) converges quickly for any $r$ if $m \geq 2$, 
but converges very slowly for $m=1$ and $(P+|Q|)r^2 << 1$.
For $m=1$ and $(P+|Q|)r^2 << 1$, an alternative form of $I_{2m}^{(2)}(r)$ 
can be found in Appendix A (eq.\ [A9]), 
which converges very quickly. 
In cartesian coordinates the deflection angle is
\begin{equation}
\vec{\alpha} \equiv \alpha_x \hat{\mbox{i}} + \alpha_y \hat{\mbox{j}} = 
(\alpha_r\cos\phi-\alpha_{\phi}\sin\phi)\hat{\mbox{i}}
 +(\alpha_r\sin\phi+\alpha_{\phi}\cos\phi)\hat{\mbox{j}}.
\end{equation}

  The determinant of the Jacobian matrix 
$\left( \frac{\partial \vec{x}_s}{\partial \vec{x}} \right)$ is
\begin{equation}
\det\left[\frac{\partial\vec{x}_s}{\partial\vec{x}}\right] 
 = \left| \begin{array}{cc} 1 - \psi,_{11}  &  -\psi,_{12}   \\
    -\psi,_{21}  &  1 - \psi,_{22} \end{array} \right|,
\end{equation}
where
\begin{eqnarray}
\psi,_{11} \equiv \frac{\partial}{\partial x}\left(\frac{\partial\psi}
{\partial x}\right) 
& = & \cos^2\phi \frac{\partial \alpha_r}{\partial r} - \cos\phi\sin\phi 
 \frac{\partial\alpha_{\phi}}{\partial r} + \frac{\sin^2\phi}{r}\alpha_r  
\nonumber \\
 & & -\frac{\sin\phi\cos\phi}{r}\frac{\partial \alpha_r}{\partial\phi} 
 +  \frac{\sin\phi\cos\phi}{r}\alpha_{\phi}+ \frac{\sin^2\phi}{r}
\frac{\partial \alpha_{\phi}}{\partial\phi},\\
\psi,_{22} \equiv \frac{\partial}{\partial y}\left(\frac{\partial\psi}
{\partial y}\right)
 & = & \sin^2\phi \frac{\partial \alpha_r}{\partial r} + \cos\phi\sin\phi
  \frac{\partial\alpha_{\phi}}{\partial r} + \frac{\cos^2\phi}{r}\alpha_r 
 \nonumber \\
 & & +\frac{\sin\phi\cos\phi}{r}\frac{\partial \alpha_r}{\partial\phi} - 
\frac{\sin\phi\cos\phi}{r}\alpha_{\phi}+ 
 \frac{\cos^2\phi}{r}\frac{\partial \alpha_{\phi}}{\partial\phi},\\
\psi,_{12}  \equiv \frac{\partial}{\partial y}\left(\frac{\partial\psi}
{\partial x}\right)
 & = & \sin\phi\cos\phi\frac{\partial\alpha_r}{\partial r} 
   -\sin^2\phi\frac{\partial\alpha_{\phi}}{\partial r}
  -\frac{\sin\phi\cos\phi}{r}\alpha_r \nonumber \\
 & & +\frac{\cos^2\phi}{r}\frac{\partial\alpha_r}{\partial \phi}  
   -\frac{\cos^2\phi}{r}\alpha_{\phi}
  -\frac{\sin\phi\cos\phi}{r}\frac{\partial \alpha_{\phi}}{\partial\phi}
 \nonumber \\
 &=&\frac{\partial}{\partial x}\left(\frac{\partial\psi}{\partial y}\right) 
\equiv \psi,_{21}. 
\end{eqnarray}
In the above expressions, it can be shown that
\begin{eqnarray}
\frac{\partial\alpha_r}{\partial r}&=& -\frac{\kappa_0}{r}I^{(0)}(r) 
  +\frac{\kappa_0}{r}I'^{(0)}(r) \nonumber \\
 & &-\frac{\kappa_0}{r} \sum_{m=1}^{\infty}\cos [m(2\phi-\delta)] 
  \left[ (2m+1)I_{2m}^{(1)}(r) + (2m-1)I_{2m}^{(2)}(r) - 4 I_{2m}^{(3)}(r) 
\right], \\
\frac{\partial\alpha_r}{r\partial\phi} &=&  
  -\frac{2\kappa_0}{r}\sum_{m=1}^{\infty} \sin [m(2\phi-\delta)]  
\left[ m I_{2m}^{(1)}(r) - m I_{2m}^{(2)}(r) \right], \\
\frac{\partial\alpha_{\phi}}{\partial r}&=& 
 -\frac{\kappa_0}{r}\sum_{m=1}^{\infty}\sin [m(2\phi-\delta)] 
 \left[(2m+1) I_{2m}^{(1)}(r) - (2m-1) I_{2m}^{(2)}(r) \right]
\end{eqnarray}
and
\begin{equation}
\frac{\partial\alpha_{\phi}}{r\partial\phi} 
 = \frac{2\kappa_0}{r}\sum_{m=1}^{\infty}\cos [m(2\phi-\delta)] 
\left[m I_{2m}^{(1)}(r) + m  I_{2m}^{(2)}(r)\right],
\end{equation}
where we have introduced two new functions: 
\begin{equation}
I'^{(0)}(r) = 2 h(r) [\varepsilon_2(r)]^{\mu} 
{_2F_1}(-\mu,-\mu;1;\varepsilon_1(r)), 
\end{equation}
and
\begin{eqnarray}
I_{2m}^{(3)}(r) & = & \left(-\frac{Q}{|Q|}\right)^m (m+\mu)\zeta_2(m,\mu)
 h(r) [\varepsilon_1(r)]^{\frac{m}{2}}
 [\varepsilon_2(r)]^{\mu} {_2F_1}(m-\mu,-\mu;m+1;\varepsilon_1(r)). 
\nonumber \\
        &  & 
\end{eqnarray}
The magnification factor is then
\begin{equation}
{\cal M} = \frac{1}{1 -(\psi,_{11} + \psi,_{22}) + \psi,_{11} \psi,_{22} -
 \psi,_{12}^2}.
\end{equation}

The coefficient functions of the deflection angle 
and magnification factor,  equations (26) - (28)
(and equations [43] and [44]), 
 contain a hypergeometric function of the same form 
${_2F_1}(j-k-\mu,-k-\mu;j+1;x)$ with $k+\mu > -1$ and $j = 0, 1, 2, ...$ 
The hypergeometric function converges for any $|x| \leq 1$. 
One can prove that the series converges rapidly 
after about the $n$-th order, where $n \approx k+\mu+1$
and higher order terms than $2n$ are almost negligible.
 Thus, one can evaluate the hypergeometic function easily 
 by truncating the series somewhere after $2n$, 
depending on the accuracy desired. Furthermore, once the value of the
function for a $k$ is known, the function values for other values of $k$
can be calculated using the Gauss recurrence relations (see, e.g.,
Abramowitz \& Stegun 1964).

For the calculation of light travel time difference (i.e.\ time delay)
 between an image pair, say at $(r_1, \phi_1)$ and $(r_2, \phi_2)$ 
on the lens plane, one needs to calculate the difference of $F(r)$ 
(eq.\ [12]) at
the two points, i.e. 
$  F(r_1) - F(r_2) =  \int_{0}^{2 \pi} d \phi' \int_{r_1}^{r_2}  
        \kappa (r',\phi') \ln r' \cdot r' dr' $ as well as the other 
coefficients
(equations [7] - [11]). For the expression used to determine
the time delay the reader is referred to SEF [equations (5.11) and (5.44)].

\section{Critical Curves and Caustics of Power-Law Lenses}
Our results can be used to see how the case of an arbitrary value of $\nu$ 
differs from the best-studied case of $\nu =2$ 
(i.e.\ the isothermal lenses studied by 
Kormann et al.\ 1994 and Kassiola \& Kovner 1993) 
in the structure of critical curves and caustics. 
 The geometry and number of images  are determined by 
 the structure of the caustics on the source plane.
For isothermal lenses, Kormann  et al. (1994) showed 
that five different types of caustics are possible in the 2-dimensional
 parameter
space spanned by axis ratio $f$ (which is related to the ellipticity
 $\epsilon$ via $\epsilon = 1 - f$) and core radius. In particular, 
the simple expressions for the deflection enabled them to map out 
the parameter space in terms of the types of caustics (see their Fig.\ 9).
Here we use examples to study the effect of varying  radial index $\nu$
as well as ellipticity and core radius 
on the behavior of critical curves and caustics.

 We compare three different values of $\nu$ including 
 the isothermal case for the same ellipticities and core radii 
 (Figures 1, 2, and 3).
This provides us with a qualitative understanding 
of the difference between shallower and steeper distributions of mass.
The case of $\epsilon=0.7$ and $r_c=0.5$ in Fig.\ 1 shows 
one liplike (i.e.\ of the diamond shape) caustic which we call 
the first caustic. 
For $\epsilon=0.5$ and $r_c=0.5$, the second caustic of a liplike 
(or pseudo-elliptic) shape appears inside the first  caustic. 
For $\epsilon=0.3$ and $r_c=0.5$, 
two cusps of the first caustic are inside the second caustic while
the other two are ``naked'' outside and 
the second caustic is of an elliptical shape.
At this stage the second caustic is called ``radial'' 
while the first caustic is called ``tangential'' (see, e.g., SEF).
 A further decrease of $\epsilon$ 
for the same value of $r_c$ causes the tangential caustic to be completely
inside the radial caustic (for $\epsilon=0.1$ and $r_c=0.5$). 
The last and trivial case is that there is no caustic.  This happens
when we sufficiently increase either $\epsilon$ or $r_c$ 
from the case of $\epsilon=0.7$ and $r_c=0.5$ in Fig.\ 1.
In the case of no caustic, multiple imaging is impossible. 
When there is one liplike caustic, three images become possible. 
When two caustics exist, up to five images are possible.

For a given value of $\nu$ the critical behavior can be summarized as follows. 
For sufficiently large ellipticity and core radius, 
no caustics exist (thereby no critical curves).
As we decrease either ellipticity or core radius, 
the first caustic comes into existence. 
A further decrease gives birth to the second caustic,
which grows as either ellipticity or core radius decreases, 
finally enclosing the tangential caustic completely 
for a sufficiently small ellipticity or core radius.

The effect of varying radial index $\nu$ on the critical behavior 
can be qualitatively understood by comparing Figures 1, 2, and 3. 
For example, for $\epsilon=0.7$ and $r_c=0.5$, 
the case $\nu = 1.5$ has one liplike caustic.
 For the same ellipticity and core radius, the isothermal case has 
two caustics. 
For $\nu = 2.5$, two cusps of the tangential caustic are inside the 
radial caustic 
(i.e.\ the second caustic is grown relative to the first caustic). 
Now the significance of $\nu$ on the critical structure is evident. 
For given ellipticity and core radius, 
a sufficiently shallow profile does not allow any caustic.
 As the profile steepens, the first  caustic will appear,
  then the second caustic, and finally
the radial caustic will enclose the tangential caustic completely.

\section{Discussion and Conclusion} 
We obtained the complete expressions for the deflection angle and 
magnification factor as (rapidly) converging series for the triaxial mass
distribution of equation (14) and for the elliptical surface density of 
equation (19). The calculation of an angle (eq.\ [34]) at a point on the 
image plane requires evaluations of three functions which are given by 
equations (26), (27) and (28). For the calculation of the magnification 
(eq.\ [45]) one needs to evaluate two additional functions 
(equations [43] and [44]). Since they converge very rapidly
all these functions can be easily evaluated. 
The hypergeometric function in equations (26), (27), (28), (43) and (44),
which is a well-behaved converging series, can be evaluated rapidly. 
For instance, one can evaluate the hypergeometric function 
by summing up less than tens of terms. However, since the hypergeometric
function should be evaluated a number of times
(namely, the function should be called many times) for the calculation of the 
deflection and magnification,  the execution speed in our approach 
depends on how the hypergeometric function is evaluated. If one evaluates
the hypergeometric function solely by summing up terms, it typically takes
about one milli-second for the calculation of the deflection and magnification
at a time on a current (SPARC20) unix machine  with a fractional uncertainty
of about $10^{-5}$ or better. For example, if the deflection angle
is one arcsecond, then the calculational error is about one hundredth 
milli-arcsecond or smaller with the execution speed specified above.

Now that our results have been presented, one can ask how well the mass model 
(eq.\ [14] or eq.\ [19]) resembles the true mass distribution
of a real lens. An application of our results to the Einstein Cross
(Q2237+0305), which is known to be lensed by an isolated barred spiral
galaxy, shows that the model is a good approximation to the true mass
distribution (Chae et al.\ 1998). In most of the other 
lensed systems external perturbations are known to be important
(Keeton et al.\ 1997). Thus, unless the external perturbations are
taken into account properly in fitting observational constraints, it is not
possible to judge whether or not the mass model is a good description of the 
true mass distribution. 
There have been numerous lens modeling examples where isothermal
models were used.\footnote{We do not list those examples here  because 
there are many.} The mass distributions of lenses in nature may differ only
slightly from isothermal distributions. However, lenses in the universe 
will not, in general, exactly mimic isothermal distributions. 
Thus, it is worthwhile to consider arbitrary values of $\nu$.
The model we considered here encompasses the 
isothermal model as a special case (i.e.\ $\nu = 2$) but allows us to vary the
parameter $\nu$ from the isothermal value. 
 
In conclusion, the following are what we consider to be the most notable 
aspects of our results.

\begin{enumerate}
\item The model we have considered represents 
the most general case of the familiar power-law mass distributions. 
It can be considered as a generalization of the isothermal model.
Hence, our results can be used to model various single mass distributions
(e.g.\ elliptical/spiral galaxies) until more physical mass models are
developed. 

\item The resulting expressions for the deflection angle and magnification 
factor are exact, and no approximations were made in any steps of the 
calculation.

\item The series expressions for the deflection angle and magnification factor 
converge rapidly and can be easily evaluated with the desired
accuracies.
\end{enumerate}

\appendix

\centerline{\bf{Appendices}}
\section{The Series Coefficients of the Deflection Angle}

We evaluate the coefficients of the deflection angle defined 
in \S 2.1, i.e., equations (7) - (11),
 for the surface mass distribution of equation (19). 
All of the odd-numbered coefficients vanish 
because the surface density
has symmetry under $\phi \rightarrow \phi + \pi$. 
Hence we set $n = 2 m$ ($m$ = 1, 2, 3,$\ldots$).

We define $g(\phi') \equiv P + Q \sin (2 \phi' + S)$ for convenience. 
Equation (8) becomes
\begin{eqnarray}
B_{2m}(r) & = & \kappa_{0} \int_{0}^{2 \pi} 
 d \phi' \cos(2m\phi')  \frac{r^{2(m+1)}}{2(m+1)} 
\frac{1}{[1+g(\phi')r^{2}]^{\mu+1}} \nonumber  \\
 &   &  \times   {_2F_1}\left(\mu+1,1;m+2;\frac{g(\phi') 
  r^{2}}{1+g(\phi') r^{2}}\right),
\end{eqnarray}
where $_2F_1(a,b;c;x)$ is the hypergeometric function.  
We write out the hypergeometric function as a series 
 and use the binomial expansion to obtain
\begin{eqnarray}
\frac{B_{2m}(r)}{r^{2m}} & = & \frac{\kappa_{0}r^2}{2(m+1)}
        \sum_{k=0}^{\infty}\frac{(\mu+1)_k}{(m+2)_k}
\int_{0}^{2 \pi}d\phi' \cos(2m\phi') 
  \frac{[r^2 g(\phi')]^k}{[1+r^2 g(\phi')]^{k+\mu+1}}   \nonumber \\
  & = & \frac{\kappa_{0}r^2}{2(m+1)}\sum_{k=0}^{\infty}
       \frac{(\mu+1)_k}{(m+2)_k } \sum_{l=0}^{k} 
       \int_{0}^{2\pi}d\phi'\cos (2m\phi') 
      \left( \begin{array}{c} k \\ l \end{array} \right) 
       \frac{[1+r^2g(\phi')]^{k-l}(-1)^{l}}{[1+r^2g(\phi')]^{k+\mu+1}}
 \nonumber \\
 & = & \frac{\kappa_{0}r^2}{2(m+1)}\sum_{k=0}^{\infty}
        \sum_{l=0}^{k}\frac{(\mu+1)_k}{(m+2)_k }
      \left( \begin{array}{c} k \\ l \end{array} \right) (-1)^{l}  \nonumber \\
 &  & \times \int_{0}^{2\pi}
     \frac{\cos 2m\phi'}{[1+r^2P+r^2Q\sin(2\phi'+S)]^{l+\mu+1}} d\phi',
\end{eqnarray}
where $(a)_b$ is the pochhammer symbol, namely $(a)_b = \Gamma(a+b)/\Gamma(a)$.
The evaluation of the integral in equation (A2) can be found in Appendix B. 
 So we have
\begin{eqnarray}
\frac{B_{2m}(r)}{r^{2m}} & = & \frac{\kappa_{0}r^2}{2(m+1)}
  \sum_{k=0}^{\infty}\sum_{l=0}^{k}\frac{(\mu+1)_k}{(m+2)_k }
  \left( \begin{array}{c} k \\ l \end{array} \right) (-1)^{l} \cos[m(\pi/2-S)] 
  \left( \frac{Q}{|Q|} \right)^m 2\pi \nonumber \\
 &  & \times \frac{[(1+r^2P)^2-(r^2Q)^2]^{-(l+\mu+1)/2}}{(-l-\mu)_m} 
  P_{l+\mu}^{m} \left(\frac{1+r^2P}{\sqrt{(1+r^2P)^2-(r^2Q)^2}}\right),
\end{eqnarray}
where $P_{l+\mu}^m(z)$ is the associated Legendre function. 
The argument $z$ $\left[\equiv (1+r^2P)/\sqrt{(1+r^2P)^2-(r^2Q)^2}\right]$ 
 $\geq 1$ and
by analytic continuation  (Gradshteyn \& Ryzhik 1994)
it can be shown that
\begin{eqnarray}
P_{l+\mu}^m(z) & = & \left( \frac{z-1}{z+1} \right)^{-\frac{m}{2}} 
\left( \frac{z+1}{2} \right)^{l+\mu} \lim_{\gamma \rightarrow -(m-1)} 
\frac{_2F_1(-l-\mu,-l-\mu-m;\gamma;\frac{z-1}{z+1})}{\Gamma(\gamma)}  
\nonumber \\
 & = & \frac{(-l-\mu)_m(-l-\mu-m)_m}{m!} \left( \frac{z-1}{z+1} 
\right)^{\frac{m}{2}} 
        \left( \frac{z+1}{2} \right)^{l+\mu} \nonumber \\
 &    & \times {_2F_1}\left(m-l-\mu,-l-\mu;m+1;\frac{z-1}{z+1}\right).
\end{eqnarray}
After substituting the above expression into equation (A3), we have
\begin{eqnarray}
\frac{B_{2m}(r)}{r^{2m}} & = & \frac{\kappa_{0}\pi r^2}{m+1}
\sum_{k=0}^{\infty}\sum_{l=0}^{k}\frac{(\mu+1)_k}{(m+2)_k }
    \left( \begin{array}{c} k \\ l \end{array} \right) (-1)^{l}
 \cos[m(\pi/2-S)]
    \left( \frac{Q}{|Q|} \right)^m \frac{(-l-\mu-m)_m}{m!} \nonumber \\
 &  & \times \frac{1}{[(1+Pr^2)^2-(Qr^2)^2]^{(l+\mu+1)/2}}
     \left( \frac{z-1}{z+1} \right)^{\frac{m}{2}} 
    \left( \frac{z+1}{2} \right)^{l+\mu} \nonumber \\
 &  & \times {_2F_1}\left(m-l-\mu,-l-\mu;m+1;\frac{z-1}{z+1}\right). 
\end{eqnarray}
Using the defined functions of $r$ (equations [30] and [31]) and 
the relation $(-l-m-\mu)_m = (-1)^m (l+\mu+1)_m$, we have
\begin{eqnarray}
\frac{B_{2m}(r)}{r^{2m}} & = & \cos [m(\pi/2-S)]  
\kappa_0 \pi r \left( -\frac{Q}{|Q|} \right)^m \frac{1}{(m+1)!}   
\frac{r}{\sqrt{(1+Pr^2)^2-(Qr^2)^2}}  [\varepsilon_1(r)]^{\frac{m}{2}} 
\nonumber \\
  &  & \times \sum_{k=0}^{\infty}\sum_{l=0}^{k} (-1)^l 
  \left( \begin{array}{c} k \\ l \end{array} \right) 
  \frac{(\mu +1)_k (l+\mu+1)_m}{(m+2)_k}  [\varepsilon_2(r)]^{l+\mu} 
\nonumber \\
 &  & \times {_2F_1}(m-l-\mu,-l-\mu;m+1;\varepsilon_1(r)).
\end{eqnarray}
As equation (A1) implies, the above expression (eq.\ [A6]) converges slowly
 when $(P-|Q|)r^2$ becomes large, i.e.\ $g(\phi')r^2/[1+g(\phi')r^2] 
\approx 1$,
while it converges rapidly for small $r$. However, equation (A1) can be 
transformed to 
an alternative form which converges rapidly for $r > 1/\sqrt{P-|Q|}$.
This alternative form is given by
\begin{eqnarray}
\frac{B_{2m}(r)}{r^{2m}} & = & \frac{\kappa_0 r^2}{2(m-\mu)} \int_0^{2\pi} 
d\phi' 
                             \frac{\cos 2m\phi'}{[1+g(\phi')r^2]^{\mu+1}}
 {_2F_1}\left(\mu+1,1;\mu+1-m;\frac{1}{1+g(\phi')r^2}\right) \nonumber \\
 &  & + \frac{\kappa_0 r^2}{2} \frac{\Gamma(m+1)\Gamma(\mu-m)}{\Gamma(\mu+1)} 
     \int_0^{2\pi} d\phi'\frac{\cos 2m\phi'}{[1+g(\phi')r^2]^{m+1}} 
\nonumber \\
 &  & \hspace{0.2in} \times {_2F_1}
            \left(m+1-\mu,m+1;m+1-\mu;\frac{1}{1+g(\phi')r^2}\right) 
\nonumber \\
 &  & (\mu \hspace{0.07in} \mbox{not integer}) \nonumber \\
 & = & -\cos(m\delta)  \kappa_0 \pi r \left(-\frac{Q}{|Q|}\right)^m h(r) 
       [\varepsilon_1(r)]^{\frac{m}{2}} 
 \frac{\Gamma(\mu-m)}{\Gamma(\mu+1)\Gamma(m+1)} \nonumber \\
 &  & \times \left\{[\varepsilon_2(r)]^{\mu} \sum_{k=0}^{\infty} 
         \frac{\Gamma(k+m+\mu+1)}{\Gamma(k-m+\mu+1)}
     [\varepsilon_2(r)]^k \right. {_2F_1}(m-k-\mu,-k-\mu;m+1;\varepsilon_1(r))
  \nonumber \\
 &  & \hspace{0.26in} \left. -[\varepsilon_2(r)]^m \sum_{k=0}^{\infty} 
      \frac{\Gamma(k+2m+1)}{\Gamma(k+1)} 
     [\varepsilon_2(r)]^k {_2F_1}(-k,-k-m;m+1;\varepsilon_1(r)) \right\}. 
\nonumber \\
  &  & (\mu \hspace{0.07in} \mbox{not integer})
\end{eqnarray}
The coefficient $C_{2m}(r)/r^{2m}$ can be calculated in a very similar way and
the resulting expressions are identical to equations (A6) and (A7), 
except that $\cos[m(\pi/2-S)]$ is replaced with $\sin[m(\pi/2-S)]$.

The coefficients $D_{2m}$ and $E_{2m}$ can also be calculated 
in a similar way. We find
\begin{eqnarray}
r^{2m} D_{2m}(r) & = & \frac{\kappa_0 r^2}{2(m+\mu)} \int_0^{2\pi}d\phi' 
\cos(2m\phi')
  \sum_{k=0}^{\infty}\frac{(\mu+1)_k}{(m+\mu+1)_k} 
    \frac{1}{[1+g(\phi')r^2]^{k+\mu+1}}   \nonumber  \\
 & = & \frac{\kappa_0 r^2}{2(m+\mu)} 
        \sum_{k=0}^{\infty}\frac{(\mu+1)_k}{(m+\mu+1)_k} \cos[m(\pi/2-S)] 
      \left( \frac{Q}{|Q|} \right)^m 2\pi \frac{(-k-\mu-m)_m}{m!}  
\nonumber  \\
 &   & \times \frac{1}{[(1+Pr^2)^2-(Qr^2)^2]^{(k+\mu+1)/2}} 
        \left( \frac{z-1}{z+1} \right)^{\frac{m}{2}} 
          \left( \frac{z+1}{2} \right)^{k+\mu} \nonumber \\
 &  & \times {_2F_1}\left(m-k-\mu,-k-\mu;m+1;\frac{z-1}{z+1}\right)  
\nonumber \\
 & = &  \cos[m(\pi/2-S)]\kappa_0 \pi r \left( -\frac{Q}{|Q|} \right)^m 
      \frac{\Gamma(m+\mu)}{\Gamma(\mu+1)\Gamma(m+1)} \nonumber \\
 &   & \times \frac{r}{\sqrt{(1+Pr^2)^2-(Qr^2)^2}} 
[\varepsilon_1(r)]^{\frac{m}{2}} [\varepsilon_2(r)]^{\mu}\nonumber \\
 &   & \times \sum_{k=0}^{\infty} [\varepsilon_2(r)]^{k}  
{_2F_1}(m-k-\mu,-k-\mu;m+1;\varepsilon_1(r)) \nonumber \\
 & = & \cos(m\delta) \kappa_0 \pi r I_{2m}^{(2)}(r)
\end{eqnarray}
and  $r^{2m} E_{2m}(r) = \sin (m\delta) \kappa_0 \pi r I_{2m}^{(2)}(r)$.

From the first line of equation (A8), it is clear that $I_{2m}^{(2)}(r)$
 converges slowly 
when $ (P+|Q|) r^2 << 1$. This is particularly true
for $m=1$. For $m \geq 2$, the convergence is still acceptable even when $ 
(P+|Q|) r^2 << 1$.
 Thus, we give an alternative form of $I_{2m}^{(2)}(r)$
for $m=1$, which converges very rapidly when $r < 1/\sqrt{P+|Q|}$:
\begin{eqnarray}
I_{2m}^{(2)}(r) & = & -\frac{P}{Q} 
   \left(1-\sqrt{1-\left(\frac{Q}{P}\right)^2}\right) r \nonumber \\
   &  & -\frac{r}{\pi} \sum_{k=1}^{\infty} \frac{(-1)^k}{k} 
    \frac{\Gamma(\mu+k+1)}{\Gamma(\mu+1)} (Pr^2)^k \sum_{l=0}^{k}
\left(\frac{2Q}{P}\right)^l \frac{[1-(-1)^l]
[\Gamma(l/2 +1)]^2}{\Gamma(k-l+1)\Gamma(l+1)\Gamma(l+2)}. \nonumber \\ 
 &  & 
\end{eqnarray}
The above expression can be obtained by dividing the integration interval of 
$r'$ 
in equation (10) into $[r,r_1]$ and $[r_1,\infty]$ where $r < r_1 < 
1/\sqrt{g(\phi')}$.

The coefficient $A_0(r)$ is obtained by substituting m=0 into the 
coefficient $B_{2m}(r)$, i.e.
\begin{eqnarray}
A_0(r) & = &  \kappa_0 \pi r\frac{r}{\sqrt{(1+Pr^2)^2-(Qr^2)^2}} 
            [\varepsilon_2(r)]^{\mu} \nonumber  \\
       &  & \times \sum_{k=0}^{\infty} \sum_{l=0}^k (-1)^l \left( 
\begin{array}{c} k \\ l 
       \end{array} \right) \frac{(\mu+1)_k}{(2)_k}[\varepsilon_2(r)]^{l} 
{_2F_1}(-l-\mu,-l-\mu;1;\varepsilon_1(r)).
\end{eqnarray}
From the hypergeometric function in equation (A1), 
it is obvious that either for a lower order of $m$ 
or a larger value of $r$, the series will converge more slowly.
 Indeed, the coefficient $A_0(r)$ converges very slowly for $r >> 1/
\sqrt{P-|Q|}$. 
 However, for large r the coefficient $A_0(r)$ can be represented by
another form which converges extremely rapidly. 
This alternative form of $A_0(r)$ can be obtained from equation (7) 
(for $\mu \neq 0$). 
Equation (7) becomes, after the integration over $r'$, 
\[
A_0(r) = \frac{\kappa_0}{2 \mu} \left[ \int_0^{2\pi} d\phi' 
\frac{1}{g(\phi')} - 
\int_0^{2\pi}d\phi' \frac{1}{g(\phi')} 
 \frac{1}{[1+r^2g(\phi')]^{\mu}} \right] \hspace{0.1 in} (\mu \neq 0).
\]
In the above expression, the second term can be easily calculated using 
$ \frac{1}{A-1} = \frac{1}{A} + \frac{1}{A^2} + \ldots $ for $A > 1$.
So we have 
\begin{eqnarray}
A_0(r) & = & \frac{\kappa_0\pi}{\mu} \frac{1}{\sqrt{P^2-Q^2}} \nonumber \\
       &   & - \frac{\kappa_0\pi}{\mu}\frac{r^2}{\sqrt{(1+Pr^2)^2-(Qr^2)^2}}
\sum_{k=0}^{\infty} 
       [\varepsilon_2(r)]^{k+\mu} {_2F_1}(-k-\mu,-k-\mu;1;\varepsilon_1(r)) 
\nonumber \\
 &  & 
\end{eqnarray}
for $\mu \neq 0$ and $r > 1/\sqrt{P-|Q|}$. 
When $\mu = 0$, for any $r$
\begin{eqnarray}
A_0(r) & = & \frac{\kappa_0\pi}{\sqrt{P^2-Q^2}} \ln \left[\frac{\sqrt{P^2-Q^2} 
           \sqrt{(P^2-Q^2)r^4 +2Pr^2 +1} +(P^2-Q^2)r^2 +P}
 {\sqrt{P^2-Q^2}+ P} \right]. \nonumber \\
 &  &
\end{eqnarray}
For a circular distribution of mass  on the lens plane (i.e.\ $Q=0$) 
all of the coefficients vanish except for  $A_0(r)$.

\section{Evaluation of Two Integrals}

For $A > |B| (\neq 0)$ we have
\begin{eqnarray}
I_c & = & \int_0^{2\pi} \frac{\cos 2m\phi}{[A+B\sin(2\phi+C)]^{\mu}} d\phi 
\nonumber \\
  & = & \int_0^{2\pi} \frac{\cos m\phi}{[A + B\sin(\phi+C)]^{\mu}} d\phi 
\nonumber \\ 
  & = & \int_{-\pi/2+C}^{3\pi/2+C} \frac{\cos(m\phi) \cos[m(\pi/2-C)] -
   \sin(m\phi) \sin[m(\pi/2-C)]} {(A+B\cos\phi)^{\mu}} d\phi \nonumber \\
  & = & \cos[m(\pi/2-C)] \int_{-\pi/2+C}^{3\pi/2+C} 
     \frac{\cos m\phi d\phi}{(A+B\cos\phi)^{\mu}} 
   - \sin[m(\pi/2-C)] \int_{-\pi/2+C}^{3\pi/2+C} 
    \frac{\sin m\phi d\phi}{(A+B\cos\phi)^{\mu}} \nonumber \\
  & = & \cos[m(\pi/2-C)] \int_{0}^{2\pi} \frac{\cos m\phi d\phi}
{(A+B\cos\phi)^{\mu}} 
   - \sin[m(\pi/2-C)] \int_{0}^{2\pi} 
    \frac{\sin m\phi d\phi}{(A+B\cos\phi)^{\mu}}. \nonumber \\
  &   & 
\end{eqnarray}
In the above equation the second term vanishes.
We use the following relationship derived from an integral representation of 
the associated Legendre function (Gradshteyn \& Ryzhik 1994)
\begin{eqnarray}
\int_0^{2\pi}\frac{\cos m\phi d\phi}{(A+B\cos \phi)^{\mu}} 
& = & 2 \int_0^{\pi}\frac{\cos m\phi d\phi}{(A+B\cos \phi)^{\mu}}  \nonumber \\
 & = & 2\pi \left(\frac{B}{|B|}\right)^m  \frac{(A^2-B^2)^{-\mu/2}}{(1-\mu)_m} 
 P_{-\mu}^m\left(\frac{A}{\sqrt{A^2-B^2}}\right). 
\end{eqnarray}
We then have
\begin{equation}
I_c = 2\pi \left(\frac{B}{|B|}\right)^m \cos[m(\pi/2-C)] 
 \frac{(A^2-B^2)^{-\mu/2}}{(1-\mu)_m} P_{-\mu}^m\left(\frac{A}
{\sqrt{A^2-B^2}}\right).
\end{equation}

Similarly, we have
\begin{eqnarray}
I_s & = & \int_0^{2\pi} \frac{\sin 2m\phi}{[A+B\sin(2\phi+C)]^{\mu}} d\phi 
\nonumber \\
    & = & 2\pi \left(\frac{B}{|B|}\right)^m \sin [m(\pi/2-C)] 
     \frac{(A^2-B^2)^{-\mu/2}}{(1-\mu)_m} P_{-\mu}^m\left(\frac{A}
{\sqrt{A^2-B^2}}\right).
\end{eqnarray}

\newpage

\begin{figure}
\centerline{\psfig{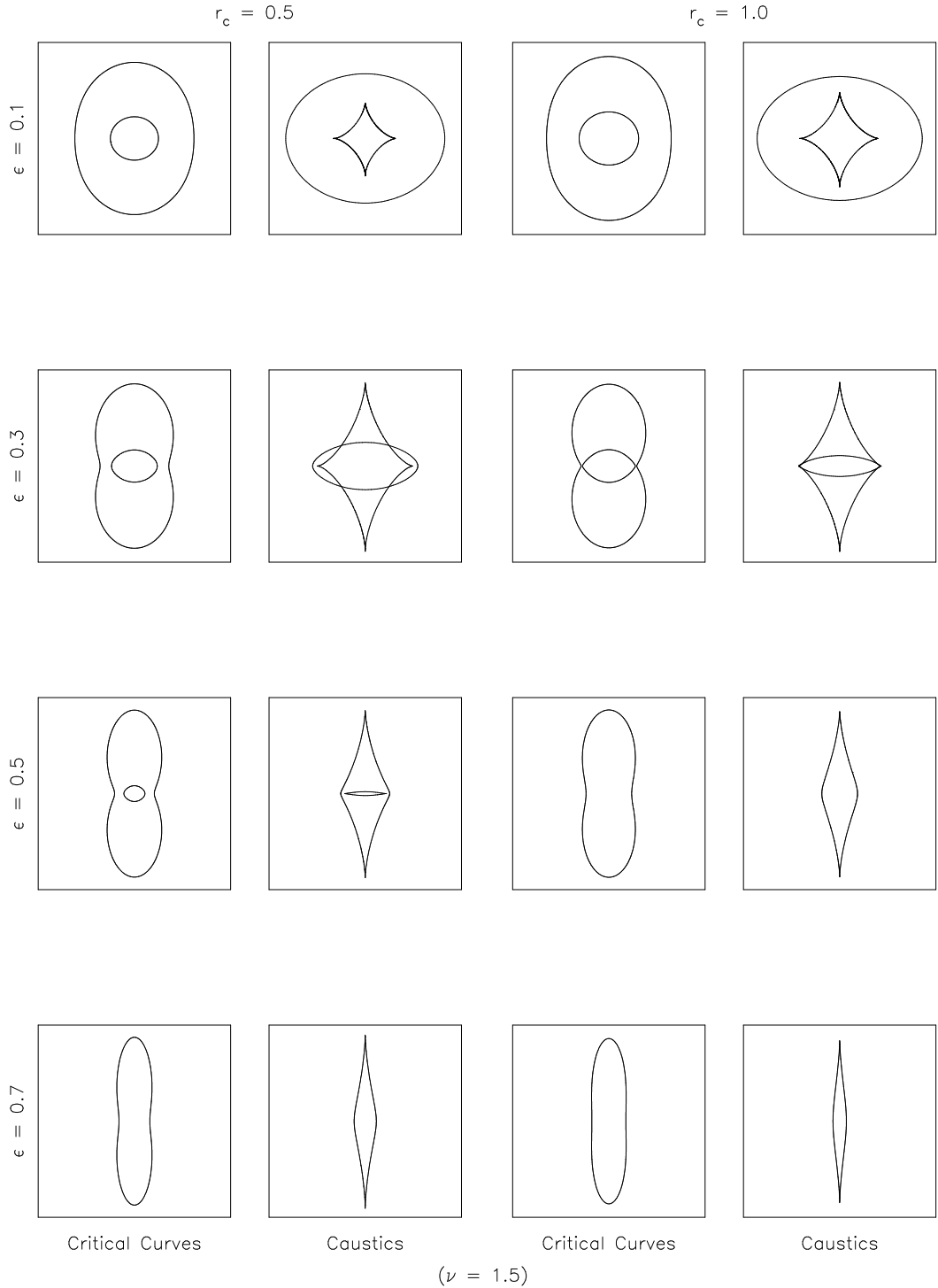}}
\caption[]{Critical Curves and Caustics for the radial index $\nu = 1.5$ 
(see \S 2.2). Two different core sizes (arbitrary scale) and four different 
ellipticities are considered.
For $r_c=0.5$, each ellipticity corresponds to a different type of caustic, 
which shows that caustic type evolves as $\epsilon$ decreases (see \S 3).
Also note that for a fixed ellipticity, caustic type evolves as $r_c$ 
decreases.}
\end{figure}

\begin{figure}
\centerline{\psfig{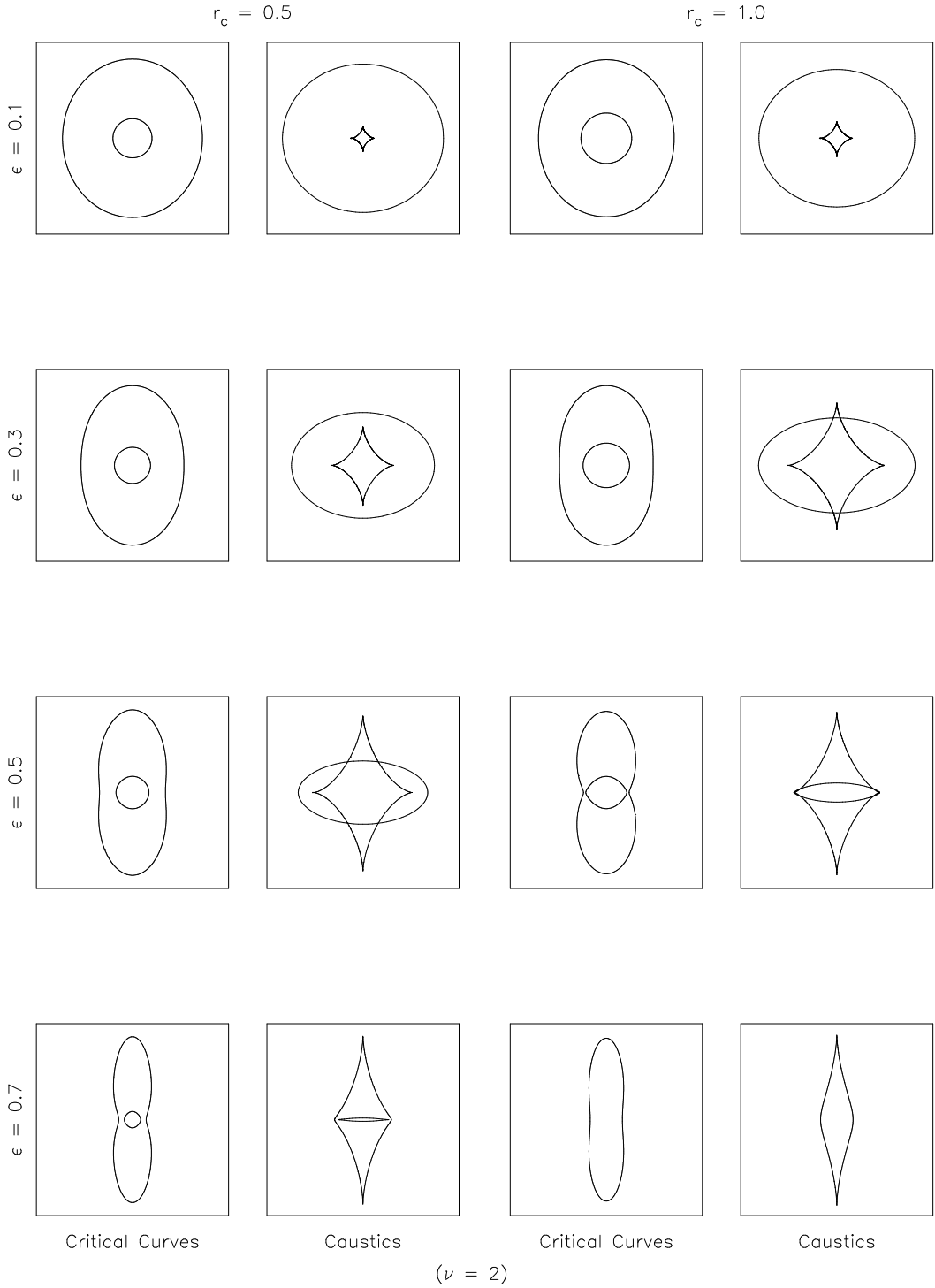}}
\caption[]{Critical Curves and Caustics for $\nu = 2$ (the
``isothermal'' distribution). 
The same core sizes and ellipticities as in Fig.\ 1 are considered.
This figure can be compared with Figures 1 and 3 
to see how caustic type varies as $\nu$ varies (see \S 3).}
\end{figure}

\begin{figure}
\centerline{\psfig{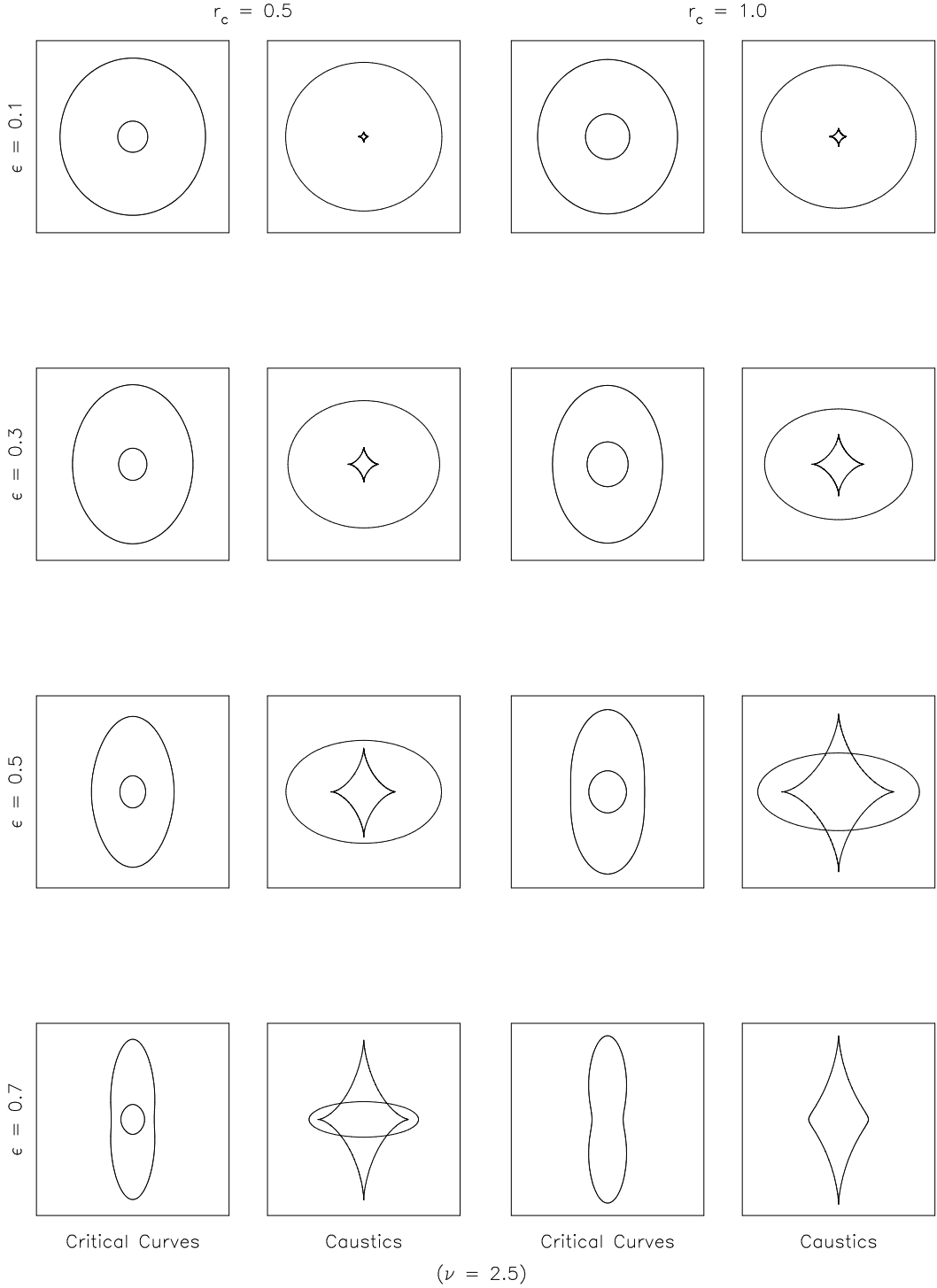}}
\caption[]{Critical Curves and Caustics for $\nu = 2.5$. 
The same core sizes and ellipticities as in Fig.\ 1 are considered.
This figure can be compared with Figures 1 and 2 to see 
how caustic type varies as $\nu$ varies (see \S 3).}
\end{figure}


\begin{references}
\reference Abramowitz, M., \& Stegun, I. 1964, Handbook of Mathematical
Functions with Formulas, Graphs, and Mathematical Tables (New York: 
Dover Publications)

\reference Blandford, R.D., \& Kochanek, C.S. 1987a, \apj, 321, 658

\reference Blandford, R.D., \& Kochanek, C.S. 1987b, in Dark Matter in the
Universe, eds.\ Bahcall, J., Piran, T., \& Weinberg, S. (Singapore: 
World Scientific) 

\reference Blandford, R.D., \& Narayan, R. 1986, \apj, 310, 568

\reference Bourassa, R.R., \& Kantowski, R. 1975, \apj, 195, 13

\reference Bourassa, R.R., Kantowski, R., Norton, T.D. 1973, \apj, 185, 747

\reference Bray, I. 1984, MNRAS, 208, 511

\reference Chae, K.-H., Turnshek, D.A., \& Khersonsky, V.K. 1998, \apj, 
           495, 609

\reference Chang, K., \& Refsdal, S. 1979, Nature, 282, 561

\reference Chang, K., \& Refsdal, S. 1984, A\&{A}, 132, 168

\reference Goldstein, H. 1980, Classical Mechanics (2nd ed.; Reading:
 Addison-Wesley)

\reference Gradshteyn, I.S., \& Ryzhik, I.M. 1994, Table of Integrals, 
           Series, and Products (5th ed.; Academic Press)

\reference Grogin, N.A., \& Narayan, R. 1996, \apj, 464, 92

\reference Kassiola, A., \& Kovner, I. 1993, \apj, 417, 450

\reference Keeton, C.R., \& Kochanek, C.S. 1997, \apj, 487, 42

\reference Keeton, C.R., Kochanek, C.S., \& Seljak, U. 1997, \apj, 482, 604

\reference Kochanek, C.S. 1991, \apj, 373, 354

\reference Kochanek, C.S., \& Blandford, R.D. 1987, \apj, 321, 676

\reference Kochanek, C.S., Blandford, R.D., Lawrence, C.R., 
           \& Narayan, R. 1989, MNRAS, 238, 43

\reference Kormann, R., Schneider, P., \& Bartelmann, M. 1994, A\&{A}, 284, 285

\reference Kovner, I. 1987a, \apj, 312, 22

\reference Kovner, I. 1987b, Nature, 325, 507

\reference Kovner, I. 1987c, \apj, 316, 52

\reference Ryden, B.S. 1992, \apj, 396, 445

\reference Schneider, P. 1985, A\&{A}, 143, 413

\reference Schneider, P., Ehlers, J., \& Falco, E.E. 1992, 
           Gravitational Lenses (New York: Springer-Verlag) (SEF)

\reference Schneider, P., \& Weiss, A. 1991, A\&{A}, 247, 269 

\reference Schramm, T. 1990, A\&{A}, 231, 19

\reference Stark, A.A. 1977, \apj, 213, 368

\reference Wambsganss, J., \&, Paczy\'{n}ski, B. 1994, \aj, 108, 1156

\end{references}
\end{document}